# An Exploration of an Early Gravity Transition in Light of Cosmological Tensions

Giampaolo Benevento, Joshua A. Kable, Graeme E. Addison, and Charles L. Bennett
Johns Hopkins University, 3400 North Charles Street, Baltimore, MD 21218, USA
*Received 2022 February 17; revised 2022 July 11; accepted 2022 July 12; published 2022 August 23*

## Abstract

We study a step-like transition in the value of the effective Planck mass (or effective gravitational constant) on cosmological scales prior to recombination. We employ cosmic microwave background, baryon acoustic oscillations, and Type Ia supernova data and find they are sufficient to strongly constrain our implementation of the effective field theory of dark energy and modified gravity, used to model the transition, to a limited parameter space. The data prefer a ∼5% shift in the value of the effective Planck mass (<10% at $2\sigma$) prior to recombination. This Transitional Planck Mass (TPM) model is free to undergo its transition at any point over multiple decades of scale factor prior to recombination, $\log_{10}(a) = -5.32^{+0.96}_{-0.72}$ (68% confidence level). This lowers the sound horizon at last scattering, which increases the Hubble constant to $71.09 \pm 0.75$ km s$^{-1}$ Mpc$^{-1}$ with a combination of local measurements as prior and to $69.22^{+0.67}_{-0.86}$ km s$^{-1}$ Mpc$^{-1}$ when the prior is excluded. The TPM model improves $\chi^2$ with respect to $\Lambda$CDM by $\Delta\chi^2 = -23.72$ with the $H_0$ prior and $\Delta\chi^2 = -4.8$ without the prior. The model allows for both $H_0 > 70$ kms$^{-1}$ Mpc$^{-1}$ and $S_8 < 0.80$ simultaneously with lower values of $S_8$ due to a reduction in the matter density $\Omega_m$ to offset the increase in $H_0$ relative to $\Lambda$CDM. While this is a particular modified gravity model, studying other variants of modified gravity may be a productive path for potentially resolving cosmological tensions while avoiding the need for a cosmological constant.

*Unified Astronomy Thesaurus concepts:* Cosmology (343); Dark energy (351); Cosmological parameters (339)

## 1. Introduction

The $\Lambda$CDM standard model of cosmology arose from a combination of diverse historical observations over many years. Cold dark matter (CDM) was introduced by Zwicky (1937), and the evidence for it was significantly strengthened by Rubin et al. (1980). CDM became a physical requirement to explain the cosmic microwave background (CMB) observations, which distinguish CDM as unaffected by radiation pressure, unlike baryons, with the measurements requiring five times more CDM than baryons (Bennett et al. 2003; Spergel et al. 2003). The surprising observational discovery of the accelerated expansion of the universe (Riess et al. 1998; Perlmutter et al. 1999) suggested the presence of an unexpected dark energy. This was soon confirmed by the WMAP CMB observations (Bennett et al. 2003; Spergel et al. 2003), which showed that the dark energy comprises ∼70% of the mass-energy budget of the flat universe, assuming a cosmological constant with the equation of state $w = -1$. Measurements of the present-day dark energy equation of state parameter, assuming standard general relativity (GR), are $w = -1.020 \pm 0.027$ (Alam et al. 2021) and $w = -1.031^{+0.030}_{-0.027}$ (Abbott et al. 2022).

Since the $\Lambda$CDM standard model of cosmology was established, the past 20 yr have seen an explosive growth of cosmological observations that have continued to support the $\Lambda$CDM model with increasingly tight parameter constraints. These recent measurement constraints on $\Lambda$CDM come from many types of observations, including the CMB (e.g., Hinshaw et al. 2013; Planck Collaboration et al. 2020a), Type Ia supernovae (SNe Ia; e.g., Scolnic et al. 2018), baryon acoustic oscillations (BAO) in the galaxy distribution (e.g., Alam et al. 2021; Eisenstein et al. 2005), and weak gravitational lensing (Heymans et al. 2021; Tröster et al. 2021; Abbott et al. 2022). Indeed, $\Lambda$CDM has survived so many observational tests that it is difficult to believe that it can be very far off from reality.

At the same time, we acknowledge that the model is ad hoc. We do not know the physical properties of the particle(s) that make up the CDM. There is an apparent cosmological constant with a particular energy density that is surprising and unexplained (Weinberg 1989). Moreover, tensions between different cosmological probes have arisen and grown in significance over the past few years.

In particular, the most direct method for measuring $H_0$, with minimal dependence on the cosmological model, is the cosmic distance ladder approach. This combines low-redshift SNe Ia in the Hubble flow ($z \lesssim 0.15$) with absolute Ia luminosity calibration, for example, from the period–luminosity relation of Cepheid variables, or tip of the red giant branch stars. Combining the SH0ES measurement with two independent and consistent determinations of $H_0$ from Pesce et al. (2020) and Blakeslee et al. (2021) obtained from the observations of masers and the IR surface brightness fluctuation in 63 galaxies results in a 5.2$\sigma$ tension with the final Planck 2018 $\Lambda$CDM value (Planck Collaboration et al. 2020a). This discrepancy remains unexplained despite extensive scrutiny. Various facets of the distance ladder have been reanalyzed, both by the SH0ES team and externally, without producing any substantial shift in the measurement (e.g., Cardona et al. 2017; Burns et al. 2018; Dhawan et al. 2018; Follin & Knox 2018; Feeney et al. 2018; Riess et al. 2019, 2020, 2021a; Dainotti et al. 2021; Perivolaropoulos & Skara 2021).

A variety of other low-redshift $H_0$ determinations have been reported recently, using methods including megamaser orbital dynamics, surface brightness fluctuations, or alternative distance ladders. Overall these measurements are consistent with one another and the SH0ES results (see Verde et al. 2019, for a review).

There is also a tension between the Planck $S_8 = \sigma_8 \sqrt{\Omega_m/0.3}$ (where $\sigma_8$ is the rms matter fluctuation averaged over a sphere of radius 8 $h^{-1}$ Mpc and $\Omega_m$ is the fractional matter density) constraint in $\Lambda$ CDM and ground-based weak lensing shear measurements from surveys including DES and KiDS. The KiDS team reported tensions with Planck at the (2.5–3.0)$\sigma$ level depending on data cuts and analysis choices (Asgari et al. 2021; Ruiz-Zapatero et al. 2021). The recently completed DES 3 yr analysis concludes that the DES results overall are consistent with Planck, but $S_8$ from the shear measurements (separate from other probes such as galaxy clustering) is lower by (2.1–2.3)$\sigma$ in the same direction as KiDS (Amon et al. 2021). Moreover, growth rate measurements from redshift space distortions also have shown a lower level of disagreement with Planck results compared to cosmic shear surveys (Efstathiou & Lemos 2018; Nunes & Vagnozzi 2021). These results suggest that the $S_8$ tension might be less severe when all possible late-time independent measurements are taken into account. Qualitatively similar tensions exist between Planck and a range of other probes of large-scale structure growth rate, including galaxy cluster abundance, and cross-correlations between galaxy surveys and CMB lensing reconstruction (e.g., McCarthy et al. 2018; Hang et al. 2021).

Even without the tensions, we would still want to know how robust $\Lambda$ CDM is or whether alternative theories could at least equally explain the data. Cosmological models that involve the presence of an additional dynamical scalar field are among the most studied extensions to the $\Lambda$ CDM model. Single scalar-field models allow for a diverse phenomenology. Quintessence (Tsujikawa 2013) and K-essence (Armendariz-Picon et al. 2001) are examples of simple scalar-field models with minimal coupling to gravity, where the equation of state (for both) and the sound speed (only for K-essence) of the scalar field are free functions of time. Despite their simplicity, such models have been shown to be able to reduce the Hubble tension with opportune choices of the equation of state parameterization.

This idea has been implemented by the class of early dark energy (EDE) models (Poulin et al. 2019). They involve the presence of an additional fluid-like component that behaves like a cosmological constant until its equation of state parameter rapidly undergoes a transition to positive values, such that its energy density dilutes faster than that of radiation. The impact of this additional component on cosmology occurs around the epoch of matter-radiation equality. This behavior can be reproduced by different theoretical models, generally characterized by a scalar field in an initial slow-roll phase, with an effective equation of state parameter $w = -1$, and a transition mechanism that allows the energy to be diluted fast enough once the field is released from the Hubble drag. The original EDE solution relied on an oscillating potential. In the acoustic dark energy (ADE; Lin et al. 2019) model, the transition in the equation of state is obtained through the conversion of the potential energy into kinetic energy, and the final value of $w$ for the scalar field is determined by the sound speed of the scalar field. Variations of the EDE scenario also include new early dark energy (Niedermann & Sloth 2021), chameleon early dark energy (Karwal et al. 2022), and $\alpha$-attractor EDE (Braglia et al. 2020), AdS-EDE (Jiang & Piao 2021). For a review and comparison of beyond-$\Lambda$CDM models proposed to address the Hubble tension (see, e.g., Schombert et al. 2020; Valentino et al. 2021).

The EDE and ADE models are examples of scalar-field models that reconcile CMB data and the local $H_0$ determination better than $\Lambda$ CDM; however, some caveats have to be considered. No substantial improvement over the standard model is found when the $H_0$ prior is not included in the analysis with Planck data (Hill et al. 2020). However, the inclusion of ACT data in the analysis leads to a preference for EDE over $\Lambda$ CDM even when the $H_0$ prior is not included (Lin et al. 2020; Hill et al. 2022). Also, to achieve the shift in $H_0$ (and also the sound horizon) while still providing an acceptable fit to CMB, BAO, and SN Ia data, these models produce shifts in the dark matter density $\omega_c$ (see Vagnozzi 2021) and scalar index, $n_s$. This leads to enhanced small-scale power in the matter power spectrum compared to $\Lambda$ CDM, effectively worsening the $S_8$ tension with weak lensing and other large-scale structure data sets (Hill et al. 2020; Ivanov et al. 2020). Moreover, the predicted EDE energy density coincidentally peaks around the epoch of matter-radiation equality, despite there being no evident physical relation between the two phenomena. This requires some fine tuning of the cosmological parameters (Sakstein & Trodden 2020).

Deviations from $\Lambda$ CDM at early times can also be produced by theories that involve modifications of gravity. Archetypal modified gravity models involve nonminimal coupling between the scalar field and the metric tensor, which induces a time variation of the effective gravitational constant appearing in the Friedmann equation. Nonminimally coupled models that induce an early-time evolution of the Planck mass offer an alternative way to modify the expansion history, possibly providing useful mechanisms to alleviate the $H_0$ tension, but at the same time, they affect the growth of perturbations, changing the shape of the CMB spectra and the value of the $S_8$ parameter inferred from Planck data. Therefore, they have the potential to solve multiple tensions at the same time. Moreover, modified gravity models can have dark energy as an emergent phenomenon, which eliminates the need for a cosmological constant. These kinds of early modified gravity models have been recently tested by different authors (Lima et al. 2016; Benevento et al. 2019; Lin et al. 2019; Peracaula et al. 2019; Ballardini et al. 2020, 2022; Ballesteros et al. 2020; Zumalacárregui 2020; Braglia et al. 2021).

In this paper we employ the effective field theory of dark energy and modified gravity (EFT), which was developed by Bloomfield et al. (2013); Bloomfield (2013); Gubitosi et al. (2013); Gleyzes et al. (2013) to test early-time variations of the effective Planck mass, induced by a nonminimal coupling of a scalar field to gravity (see also, e.g., Frusciante & Perenon 2020, for a review). The EFT formalism allows us to describe departures from general relativity in terms of a set of time-dependent functions which have direct connections with observable properties. We adopt a pure EFT approach, where we do not specify the explicit form of the Lagrangian operators in terms of the scalar field and its time derivative, but rather we fix the time dependence of the EFT functions. Previous works, like Raveri (2020), have adopted a similar approach to perform a nonparametric reconstruction of the EFT functions. Here we are interested in exploring specific features in the temporal evolution of the EFT functions, which makes a parametric description the most natural choice.

In particular, we study the simple but meaningful case of the Planck mass undergoing a step-like transition, described by an error function of the number of e-folds. In the late universe, the scalar field in this model takes on the role of dark energy and has an equation of state parameter that is not fixed to −1 but rather constrained by data. This eliminates the need for a cosmological constant to be implemented by hand as in the Λ CDM scenario. For the present analysis, we use a combination of CMB, BAO, and SN Ia data to constrain the model. Additionally, we perform tests both with and without a prior on $H_0$ resulting from the latest constraints from the SH0ES team Riess et al. (2021b). We use the natural units where the speed of light and the reduced Planck constant are fixed to 1, and the reduced Planck mass is $m_0 = (8\pi G)^{-1/2}$.

In Section 2, we describe the definition of the model in terms of EFT functions. In Section 3, we outline the data sets used in the likelihood for the Markov Chain Monte Carlo (MCMC) analysis. In Section 4, we discuss the model phenomenology and present the results. In Section 5, we provide conclusions.

## 2. Transitional Planck Mass Model

We seek a model that has a transition of the value of the gravitational constant (or equivalently, the Planck mass) on cosmological scales, prior to recombination, which we call the Transitional Planck Mass (TPM) model. Transitions in the Planck mass are common in the context of modified gravity, in particular nonminimally coupled models such as $f(R)$ (Sotiriou & Faraoni 2010), Jordan-Brans-Dicke (Brans & Dicke 1961), DGP (Deffayet 2002), and K-mouflage (Brax & Valageas 2014) predict an evolving Planck mass.

The effective Planck mass on cosmological scales in nonminimally coupled models is not necessarily equal to our solar system value, as discussed, for example, by Bellini & Sawicki (2014). At the linear perturbation level, the Newton's constant regulating the gravitational interaction between test masses receives additional contributions and can become scale dependent, as we show for the TPM model in Section 4.1. At small astrophysical scales, the scalar-field dynamics are dominated by nonlinearities, which should screen the effect of modified gravity and allow the model to match solar system constraints. These kinds of screening mechanisms are a common feature of modified gravity models (see, e.g., Koyama 2016; Lombriser 2016; Brax et al. 2022).

We will now define the TPM model in terms of the functions of the EFT. In Section 2.1, we review the definition of the EFT. In Section 2.2, we state the assumptions of the TPM model and define the model in terms of the EFT functions. In Section 2.3, we discuss the mapping of the TPM model into the Horndeski action and stability conditions that the model must satisfy.

### 2.1. Effective Field Theory of Dark Energy and Modified Gravity

The EFT is built by selecting all operators consistent with time-dependent spatial diffeomorphism invariance up to a certain order in perturbation theory. At each order in perturbations, there is a finite number of such operators that enter the action, and each one is multiplied by a time-dependent coefficient. The most general action compatible with the above requirement up to quadratic order is:

$$S = \int d^4x \sqrt{-g} \left\{ \frac{m_0^2}{2}[1 + \Omega(t)]R + \Lambda(t) - c(t)\delta g^{00} \right.$$
$$+ \frac{M_2^4(t)}{2}(\delta g^{00})^2 - \frac{\bar{M}_1^3(t)}{2}\delta g^{00}\delta K_\mu^\mu - \frac{\bar{M}_2^2(t)}{2}(\delta K_\mu^\mu)^2$$
$$- \frac{\bar{M}_3^2(t)}{2}\delta K_\nu^\mu \delta K_\mu^\nu + \frac{\hat{M}^2(t)}{2}\delta g^{00}\delta R^{(3)}$$
$$+ m_2^2(t)(g^{\mu\nu} + n^\mu n^\nu)\partial_\mu(g^{00})\partial_\nu(g^{00}) \Big\}$$
$$\left. + S_M[g_{\mu\nu}, \psi_m^{(i)}] \right.$$
(1)

where $n_\mu = -\partial_\mu \phi / \sqrt{-\partial_\mu \phi \partial^\mu \phi}$ is the unit vector perpendicular to the time slicing, $R$ is the four-dimensional Ricci scalar, $\delta g^{00}$, $\delta K_\mu^\nu$, $\delta K_\mu^\mu$ and $\delta R^{(3)}$ are respectively the perturbations of the upper time-time component of the metric, the extrinsic curvature and its trace, and the three-dimensional spatial Ricci scalar. The factors $\{\Omega, \Lambda, c, M_2^4, \bar{M}_1^3, \bar{M}_2^2, \bar{M}_3^2, m_2^2, \hat{M}^2\}$ are unknown functions of time, and we refer to them as EFT functions. In particular, $\{\Omega, \Lambda, c\}$ are the only functions contributing both to the dynamics of the background cosmology and to the perturbations, while the others contribute only to the evolution of cosmological perturbations, and express different phenomenology.[1] The EFT functions $\Lambda$ and $c$ contribute to the evolving equation of the state of the fluid associated with the scalar field that is modifying gravity, while the $\Omega$ function can be interpreted as a rescaling of the reduced Planck mass, $m_0$, on the cosmological background. In particular, the effective Planck mass, $M_\star$, on large scales at the background level, is given by $M_\star^2 = m_0^2(1 + \Omega)$. The EFT function $M_2^4$ is sourced by a nonstandard kinetic term of the scalar field, $\bar{M}_1^3$ is sourced by the mixing between the metric and the scalar field, $\bar{M}_2^2$ is related to a nonstandard speed of gravitational waves, while all the other functions describe deviations from Horndeski models (Horndeski 1974), i.e., models where the scalar-field equation of motion is not second order.

### 2.2. Parameterizing the EFT Functions

The TPM model is defined by a direct choice of the operators of the EFT action in Equation (1). We restrict our choice to the set of EFT functions that lead to the second-order equation of motion, i.e., we select models that fall within the Horndeski class. Moreover, we impose that the speed of gravitational waves is constant and equal to the speed of light, in agreement with the constraint found by Abbott et al. (2017). After these assumptions we are left with the following subset of EFT functions $\{\Omega, \Lambda, c, M_2^4, \bar{M}_1^3\}$.

In the EFT, the background expansion history is described by four functions of time: $\Omega$, $\Lambda$, $c$, and $H$, which is the standard time-dependent Hubble expansion rate. The equations of motion at the background level fix two of them, leaving us with the freedom of choosing the remaining ones.

---

[1] Note that $\Omega$, $c$, and $\Lambda$ in this context are not to be confused with the fractional energy density of the universe, the speed of light, or the cosmological constant as is typically used in cosmology. Hereafter, we adopt the convention of using $\Omega$, $c$, and $\Lambda$ to refer to the EFT background functions for consistency with previous works using the EFT language.

Although the most effective attempts of solutions to the Hubble tension are those that modify the cosmic expansion history at early epochs (see, e.g., Knox & Millea 2020; Schöneberg et al. 2021), leading to a reduced sound horizon at recombination, it was shown that this change alone did not fully resolve the $H_0$ tension while remaining consistent with other cosmological data sets (Jedamzik et al. 2021).

We seek possible solutions to the Hubble tension where the change in the background expansion at early times is driven by $\Omega$, but we also allow a late-time variability in the equation of state, determined by the EFT function $c$.

In particular, we study the simple case in which $\Omega$, and correspondingly the cosmological value of the effective Planck mass, $M_\star$, undergoes a smooth transition between two values, prior to recombination. This assumption is reminiscent of what happens to $w_{\rm EDE}$ in EDE models. Similar transitions in the Planck Mass have been also considered during inflation (Ashoorioon et al. 2014), during the matter-dominated epoch (Sakr & Sapone 2022), and at very low redshift (Marra & Perivolaropoulos 2021; Alestas et al. 2022). More complex parameterizations are possible; for instance, one could explore the case of a double transition in $M_\star$ at both early and late times. While this could be an interesting choice for future work, we notice that more complex choices of the $\Omega$ function would inevitably increase the number of free parameters of the model, and would be constrained by stability conditions. We therefore express the $\Omega$ function with a step-like function ERF(x), with $x = \log(a)$ such that

$$\Omega(x) = \frac{\Omega_0}{2}\left(1 - ERF\left(\frac{(\mu - x)}{\sqrt{2\pi}\,\sigma}\right)\right) \quad (2)$$

$$\Omega'(x) = \Omega_0 \frac{\exp\frac{-(x-\mu)^2}{2\sigma^2}}{\sqrt{2\pi}\,\sigma} \quad (3)$$

where $\mu = \log(a_T)$ and $a_T = 10^{x_T}$ determines the transition scale factor,[2] $\sigma$ sets the width of transition, and $\Omega_0$ sets its amplitude, assuming an initial value $\Omega(x_i) = 0$. This choice makes the effective Planck mass today different from the local value on solar system scales, requiring a screening mechanism to be effective. We leave to a future exploration the study of TPM properties on nonlinear scales, but we emphasize that the general features of the pure EFT model we consider could in principle be reproduced by different scalar-field theories, with different screening mechanisms. Normalizing $\Omega$ such that general relativity is recovered at early times before the transition helps guarantee that big bang Nucleosynthesis (BBN) constraints are satisfied by the TPM model, as we discuss in Section 3.2.

The remaining freedom at the background level is fixed by the choice of $c$. We require the $c$ function to be a constant

$$\frac{c}{3H_0^2 m_0^2} = c_0, \quad (4)$$

at all times. This choice guarantees that the early-time contribution of $c$ to the energetic balance of the universe is completely negligible (being suppressed by a factor $1/H^2$). Also it introduces only one additional parameter for the parameterization of $c$. We also fix the remaining EFT functions $M_2^4$ and $\bar{M}_1^3$ as follows:

$$\frac{M_2^4}{3H_0^2 m_0^2} = -c_0, \quad (5)$$

$$\frac{\bar{M}_1^3}{3H_0^2 m_0^2} = \frac{2c_0}{H}. \quad (6)$$

The above choice guarantees that the effect of $M_2^4$ and $\bar{M}_1^3$ on the linear cosmological perturbations is suppressed by a factor $1/H^2$, as we will show.

The $\Lambda$ and $H$ function are determined by the solution of the Friedmann equations:

$$H^2 = \frac{1}{1 + \Omega + \Omega'}\left(\frac{\rho_{m,rad}}{3m_0^2} - \frac{\Lambda}{3m_0^2} + \frac{2c}{3m_0^2}\right) \quad (7)$$

$$H' = -\frac{H^2 m_0^2(3 + 3\Omega + 2\Omega' + \Omega'') + \Lambda + P_{m,rad}}{H m_0^2 (2 + 2\Omega + \Omega')} \quad (8)$$

where prime denotes derivatives that are taken with respect to the variable $x \equiv \log(a)$. We follow this same convention throughout the paper, unless otherwise specified.

### 2.3. Connection to Horndeski Gravity

The TPM model can be expressed as a subclass of Hornedski models, specified by the action

$$\mathcal{S} = \int d^4x\sqrt{-g}\left[\frac{m_0^2}{2}G_4(\phi)R + G_2(\phi, X) + G_3(X)\,\Box\,\phi\right] + \mathcal{S}_m(\psi_m, g_{\mu\nu}), \quad (9)$$

where $X = \nabla^\mu\phi\nabla_\mu\phi$. In the usual approach, the form of the $G_i$ functions are specified as analytic functions of $\phi$ and $X$. In the pure EFT approach, the $G_i$ can be determined as a function of time after the solution of Equations (7)–(8), using the mapping rules presented by, for example, Frusciante et al. (2016).

A special case is represented by the choice of $c_0 = 0$, which corresponds to a standard $f(R)$ model, which we refer to as TPM f(R), after the following identification

$$\Lambda = f(R) - R\frac{df}{dR}, \quad \Omega = \frac{df}{dR}. \quad (10)$$

The expression of $G_i(\phi, X)$ in then given by

$$G_2(\phi, X) = -f(R) + R\frac{df}{dR}, \quad (11)$$

$$G_3(X) = 0, \quad (12)$$

$$G_4(\phi) = \phi, \quad (13)$$

$$\phi = 1 + \frac{df}{dR}, \quad R = 6(2H^2 + \dot{H}). \quad (14)$$

The general TPM model can be easily recast in terms of Horndeski $\alpha$ functions (Bellini & Sawicki 2014). This formalism is based on a combination of Horndenski operators forming a set of dimensionless functions that depend on time and represent distinct physical properties of the theory. The $\alpha$ functions regulate the evolution of linear perturbations in

---

[2] Note that we define $x$ as the number of e-folds, i.e., using the natural logarithm, but $x_T$ is defined in terms of the common $\log_{10}$ of the scale factor, for the readability of the final results.

Horndeski theories. For the TPM model, we can identify the following correspondences:[3]

$$\alpha_M = \frac{\Omega'}{1+\Omega}, \quad (15)$$

$$\alpha_B = \frac{\Omega'}{2(1+\Omega)} + \frac{c_0}{H^2(1+\Omega)}, \quad (16)$$

$$\alpha_K = -\frac{2c_0}{H^2(1+\Omega)}. \quad (17)$$

The nonzero $\alpha_M$ modifies the growth of structure, sourcing anisotropic stress. The $\alpha_B$ function is sourced by a kinetic mixing between gravitational and scalar degrees of freedom when the second time derivatives of the gravitational potential enter the equation of the scalar field and vice-versa. This kinetic mixing between the metric and scalar degrees of freedom is a characteristic property of theories with nonminimal coupling to gravity. The $\alpha_K$ function modulates the speed of sound for scalar perturbations and enters the condition for the absence of a scalar ghost.

In the assumption of an early-time transition ($a_t = 10^{x_T} \ll 1$ with a transition width, $\sigma \sim 1$), two different regimes are clearly separated. At early times, relatively close to the transition, we have $\Omega' \gg c_0/H^2$, and the TPM model behaves like a pure f(R) model, with $\alpha_M = 2\alpha_B$, $\alpha_K = 0$. At late times and far from the transition, the TPM model enters a regime in which $\alpha_M = 0$ and $\alpha_K \propto \alpha_B \propto c_0/H^2$, which resembles a kinetic gravity braiding model (Deffayet et al. 2010). In Section 4.1 we show that these two regimes determine a different effect on the growth of matter perturbations.

Throughout the analysis we impose the requirement of having a theory free from ghost and gradient instabilities. The no-ghost condition can be expressed as $\alpha_K + \frac{3}{2}\alpha_B^2 > 0$, and we see from Equation (17) that this is always satisfied for $c_0 < 0$, which implies $\alpha_K > 0$. The no-gradient instability condition is expressed by the requirement of having a positive speed of propagation of scalar perturbations squared $c_s^2 > 0$. The $c_s^2$ can be expressed in terms of the $\alpha$ functions defined above for TPM as

$$H^2\left(\alpha_K + \frac{3}{2}\alpha_B^2\right)c_s^2 = -(2+2\alpha_B)[\dot{H} + H^2(\alpha_B - \alpha_M)]$$
$$- 2H\dot{\alpha}_B - (\rho_{m,\mathrm{rad}} + P_{m,\mathrm{rad}})/(1+\Omega). \quad (18)$$

The stability conditions are satisfied throughout the cosmic evolution by the choice $\Omega_0 < 0$ and $c_0 < 0$.

The $\Lambda$CDM model is recovered in the limit that $\Omega_0 \to 0$ and $c_0 \to 0$, in which the EFT functions $\Omega$, $c$, $M_2^4$, and $\bar{M}_1^3$ vanish, and the $\Lambda$ becomes the cosmological constant.

## 3. Data Analysis

### 3.1. Numerical Implementation

In this subsection, we outline our use of MCMC to constrain the TPM model parameters. In recent years, EFTCAMB[4] (Hu et al. 2014; Raveri et al. 2014), an extended version of the CAMB (Lewis et al. 2000) code, was developed to include single field dark energy and modified gravity theories using the EFT formalism. The EFTCosmoMC code, an extended version of CosmoMC (Lewis & Bridle 2002), is interfaced with EFTCAMB to run MCMC and explore the EFT parameter space. We modified the EFTCAMB code to include the TPM model.

---

[3] We follow the EFTCAMB convention (Frusciante et al. 2016; Hu et al. 2014).
[4] https://github.com/EFTCAMB/EFTCAMB

The code solves numerically the two Friedmann Equations (7)–(8) to determine the background evolution of the model and the EFT functions, using the definitions in Section 2.2. We solve the equations backward in time, starting at a scale factor $a = 1$, and we impose the initial condition $H(a = 1) = H_0$. After the background equations are solved, EFTCAMB solves the linearly-perturbed Einstein–Boltzmann equations, which can be found in Hu et al. (2014).

For the data analysis we ran MCMCs using EFTCosmoMC. We used the Gelman–Rubin convergence statistic $R - 1 = 0.02$ for the least constrained parameter to determine when the chains have converged (Gelman & Rubin 1992).

After running MCMCs we maximized the posterior distribution, using the BOBYQA routine (Powell 2009) embedded in the EFTCosmoMC code. We discuss the best-fit $\chi^2$ values thus obtained in the next section, and use them for the statistical comparison of the TPM and $\Lambda$CDM models and to produce the plots in Section 2. We estimated the support for TPM over $\Lambda$CDM using the Bayesian evidence factor ($\log_{10} B$), which takes into account the full posterior probability along with the number of free parameters, as discussed in Szydłowski et al. (2015). We computed evidence factors using the code provided by Heavens et al. (2017). A positive value of $\Delta \log_{10} B$ indicates a statistical preference for the TPM model, and a strong preference is defined for $\Delta \log_{10} B > 2$.

### 3.2. Observational Data

In this subsection we provide the data sets we used to constrain the TPM model. Our baseline data sets include the Planck CMB likelihood, various BAO measurements, and a complete catalog of type Ia supernovae. In detail:

1. For CMB, we use the 2018 Plik Lite TTTEEE + Low ell TT + Low E likelihoods as well as the Planck 2018 Lensing likelihood (Planck Collaboration et al. 2020b).
2. For BAO, we use the completed SDSS-III Baryon Oscillation Spectroscopic Survey (BOSS) survey DR 12 (Alam et al. 2017), the SDSS Main Galaxy Sample (Ross et al. 2015), and the 6dFGS survey (Beutler et al. 2011).
3. For Type Ia supernovae, we use the 2018 Pantheon compilation (Scolnic et al. 2018).
4. In cases where we include an $H_0$ prior, we use $72.61 \pm 0.89$, which is obtained by combining the independent results of Riess et al. (2021b) with Pesce et al. (2020) and Blakeslee et al. (2021).

Hereafter, we refer to the likelihood combination above excluding the $H_0$ prior from local measurements as the Baseline likelihood. When the $H_0$ prior from local measurements is included, we refer to this as Baseline + $H_0$. We stress that the TPM model does not show very low redshift ($z \ll 0.1$) variation in the equation of state. Thus, a direct inclusion of an $H_0$ prior in the data set is, in this case, practically equivalent to the use of Type Ia supernovae calibrated with local measurements, as discussed in Benevento et al. (2020) and shown in Dhawan et al. (2020).

In general, the best-fit TPM model prefers an enhanced CMB lensing power spectrum relative to the best-fit Λ CDM CMB lensing power spectrum. This is most apparent at higher multipole moments. For the Planck lensing likelihood in our analysis, we use the suggested multipole cut given by $8 \leqslant L \leqslant 400$. We tested the best-fit parameterization for the TPM model against both the Planck 2018 aggressive multipole cut $8 \leqslant L \leqslant 2048$ (Planck Collaboration et al. 2020c) as well as against the SPTpol 2019 lensing likelihood (Wu et al. 2019) and found that our conclusions are robust to the choice of CMB lensing likelihood. Quantitatively, we performed a simple $\chi^2$ test using the Planck aggressive multipole cut likelihood for both the best-fit TPM and Λ CDM models, finding that Λ CDM fits slightly better with $\Delta\chi^2 = 1.5$. Future CMB lensing data will likely be a useful method of distinguishing between the Λ CDM and TPM models.

In our analysis, we do not include measurements of primordial light element abundances from BBN. While BBN is sensitive to the early universe expansion rate (see, e.g., Alvey et al. 2020), as long as the Planck mass on cosmological scales during BBN is equivalent to the Λ CDM case, the Hubble expansion rate during BBN will also be equivalent to the Λ CDM case. For the TPM model, we assume that the transition in $\Omega_0$ happens after BBN and that the cosmological value of the Planck mass during BBN is equivalent to the local value today. Moreover, we find that the TPM model prefers small shifts in the value of $\Omega_b h^2$ relative to Λ CDM when both are fit to the Baseline + $H_0$ likelihood. These two facts imply that the TPM model predicts similar values for the primordial light element abundances as Λ CDM. We checked this and found that the shifts in the primordial light element abundances are small relative to the current constraints.

Including weak lensing data would provide insight into whether the TPM model is resolving the clustering or $S_8$ tension. The TPM model does lead to an enhancement of matter perturbation modes on small scales $k \gtrsim 10^{-1}$ h Mpc$^{-1}$, as we will show in Section 4, which may be disfavored by weak lensing data. However, this enhancement affects scales that are nonlinear, meaning that to include weak lensing data would require additional code to handle nonlinear effects. We plan to perform this test in the future.

## 4. Results

We have fit the TPM and TPM f(R) models to a data set that combines CMB anisotropy and lensing data, BAO surveys, Type Ia supernovae, and a prior on $H_0$ from local measurements, which we refer to as the Baseline + $H_0$ likelihood. We find that the model provides a substantial improvement over the Λ CDM fit to the same data. The constraints on the full set of independent cosmological parameters for the TPM, TPM f(R), and Λ CDM models are summarized in Tables 1 and 2. The constraints in Table 1 are obtained from the Baseline + $H_0$ likelihood, while the constraints in Table 2 do not include local determinations of $H_0$. In Section 4.1, we discuss the phenomenology of the model, and in Section 4.2 we discuss the parameter constraints resulting from the MCMC fits to the Baseline + $H_0$ and Baseline likelihoods.

### 4.1. Model Phenomenology

The background evolution of the TPM model can be described through an effective fluid with pressure and density defined in the following. Rewriting the Equations 7–8 as

$$H^2 = \frac{1}{3m_0^2}(\rho_{m,rad} + \rho_{\text{TPM}}), \quad (19)$$

$$H' = -\frac{3H^2 m_0^2 + P_{\text{TPM}} + P_{m,rad}}{2H m_0^2}, \quad (20)$$

where we can identify the following correspondences

$$\rho_{\text{TPM}} = 2c - \Lambda - 3m_0^2 H^2(\Omega' + \Omega), \quad (21)$$

$$P_{\text{TPM}} = \Lambda + H^2 m_0^2 \left[\Omega'' + \Omega'\left(\frac{H'}{H} + 2\right) + \Omega\left(2\frac{H'}{H} + 3\right)\right]. \quad (22)$$

The effective TPM fluid density and pressure satisfy a standard continuity equation $\rho_{\text{TPM}}' = -3(P_{\text{TPM}} + \rho_{\text{TPM}})$, and deviates from the usual ΛCDM background. The evolution of the fluid's equation of state is $w_{\text{TPM}} \equiv P_{\text{TPM}}/\rho_{\text{TPM}}$, which is shown in Figure 1 together with the $\Omega$ function and the quantity $\Delta H(a)/H_{\Lambda\text{CDM}} \equiv (H_{\text{TPM}} - H_{\Lambda\text{CDM}})/H_{\Lambda\text{CDM}}$, for different choices of the parameters $\Omega_0$, $x_T$, $\sigma$, and $c_0$. For typical values of $\sigma \sim 1$, the $\Omega$ function evolves during a small fraction of the expansion history, as seen in the first panel of Figure 1. The epoch of transition is set by the $x_T$ parameter.

After the transition when $x > \log(a_T) + \sigma$, we can make the simplifying assumption $\Omega \to \Omega_0$, $\Omega' = \Omega'' = 0$, and the equation of state takes the form

$$w_{\text{TPM}} = \frac{\Lambda - \Omega_0 P_{m,\text{rad}}}{2c - \Lambda - \Omega_0 \rho_{m,\text{rad}}}. \quad (23)$$

From Equation 23 we see that the TPM equation of state resembles that of the dominant component of the energy balance whenever $\Omega_0 \rho_{m,\text{rad}} \gg 2c - \Lambda$, in particular when approaching the solution $w_{\text{TPM}} = 0$ during the matter-dominated epoch. For typical values of the TPM parameters, the above condition is satisfied until late times ($a \sim 1$), where $w_{\text{TPM}} = \Lambda/(2c - \Lambda) \simeq -1$, recovering negative values required to act as dark energy. During this phase, the $c_0$ parameter plays an important role in changing the value of $w_{\text{TPM}}$. Lower values of $c_0$ give a lower $w_{\text{TPM}}$ during the dark energy-dominated epoch, for a scale factor $a \sim 1$. Values of $w_{\text{TPM}} < -1$ at late times, which have been shown to help reduce the $H_0$ tension (Lee et al. 2022), are allowed and compatible with stability requirements. The role of the $\Omega_0$ parameter is that of determining the value of $\Omega(a)$ after the transition. When $\Omega_0$ is closer to 0, the $H(a)$ for TPM approaches the Λ CDM solution after the transition.

At very high redshift, before the transition, a similar argument applies, and the equation of state approaches −1 for $x \ll \log(a_T)$. During the transition, for $x \sim \log(a_T)$, the $w_{\text{TPM}}$ function interpolates between the positive value $w_{m,rad}$ and −1. For $x < \log(a_T)$ the non-negligible contribution of $\Omega'$, $\Omega''$ will push the equation of state to negative values. The parameter $\sigma$ plays a significant role close to the transition because it determines the rapidity of the evolution of $\Omega$, $w_{\text{TPM}}$, and $\Delta H/H_{\Lambda\text{CDM}}$.

We can understand the qualitative behavior of the fluid equation of state parameter in the TPM model, by considering that this model assumes a flat universe at all times. Changing

**Table 1**
Maximum Likelihood (ML) Parameter Values and, in Parentheses, Mean plus 68% Confidence Level (CL) Bounds, for the $\Lambda$ CDM, TPM f(R), and TPM Models, Using Our Baseline + $H_0$ Data Set

| | Fit To Baseline + $H_0$ Likelihood | | |
|---|---|---|---|
| Model | $\Lambda$CDM | TPM $f(R)$ | TPM |
| $100\theta_{MC}$ | 1.04131 (1.04126 ± 0.00029) | 1.04195 (1.04201 ± 0.00035) | 1.04154 (1.04163 ± 0.00037) |
| $\Omega_b h^2$ | 0.02257 (0.02260 ± 0.00013) | 0.02267 (0.02275 ± 0.00015) | 0.022649 (0.02261 ± 0.00015) |
| $\Omega_c h^2$ | 0.11777 (0.11752 ± 0.00087) | 0.1181 ($0.1184^{+0.0011}_{-0.00095}$) | 0.1191 (0.1190 ± 0.0010) |
| $\tau$ | 0.0621 ($0.0631^{+0.0070}_{-0.0084}$) | 0.0602 ($0.0616^{+0.0072}_{-0.0082}$) | 0.054 (0.0542 ± 0.0076) |
| $\ln(10^{10} A_s)$ | 3.058 ($3.058^{+0.014}_{-0.016}$) | 3.060 (3.063 ± 0.016) | 3.049 (3.046 ± 0.016) |
| $n_s$ | 0.9698 (0.9712 ± 0.0037) | 0.9805 (0.9808 ± 0.0046) | 0.9772 (0.9751 ± 0.0047) |
| $\Omega_0$ | ⋯ | −0.063 (−0.065 ± 0.018) | −0.045 (−0.050 ± 0.019) |
| $x_T$ | ⋯ | −4.26 ($-5.30^{+1.0}_{-0.59}$) | −4.29 ($-5.32^{+0.96}_{-0.72}$) |
| $x_T$ (95% CL) | | $-5.30^{+1.2}_{-1.5}$ | $-5.3^{+1.3}_{-1.6}$ |
| $\sigma$ | ⋯ | 0.82 ($0.97^{+0.27}_{-0.83}$) | 0.88 ($1.04^{+0.34}_{-0.88}$) |
| $\sigma$ (95% CL) | | <1.95 | <2.12 |
| $c_0$ | ⋯ | 0 (fixed) | −0.0176 ($-0.0148^{+0.0025}_{-0.0050}$) |
| $c_0$ (95% CL) | | | $-0.0148^{+0.0085}_{-0.0066}$ |
| $w_{TPM,0}$ | ⋯ | −0.9735 (−0.9728 ± 0.0072) | −1.0249 ($-1.025^{+0.013}_{-0.020}$) |
| $H_0$ | 68.44 (68.54 ± 0.40) | 70.80 (70.90 ± 0.76) | 71.38 (71.09 ± 0.75) |
| $\sigma_8$ | 0.810 ($0.8090^{+0.0059}_{-0.0066}$) | 0.868 ($0.8386^{+0.0068}_{-0.032}$) | 0.854 ($0.8531^{+0.0029}_{-0.033}$) |
| $S_8$ | 0.811 (0.8086 ± 0.0099) | 0.842 ($0.813^{+0.010}_{-0.033}$) | 0.825 ($0.8264^{+0.0065}_{-0.034}$) |
| $\Delta\chi^2_{CMB}$ | 0 | 1.8 | −6.29 |
| $\Delta\chi^2_{CMB\,lensing}$ | 0 | 0.01 | −0.57 |
| $\Delta\chi^2_{BAO}$ | 0 | −0.07 | 1.67 |
| $\Delta\chi^2_{Pantheon}$ | 0 | 0.0 | 1.35 |
| $\Delta\chi^2_{H_0}$ | 0 | −17.8 | −20.06 |
| $\Delta\chi^2_{tot}$ | 0 | −16.68 | −23.72 |
| $\Delta\log_{10} B$ | 0 | 5.69 | 4.93 |

**Note.** Mean plus 95% CL bounds are also shown for selected parameters.

the Planck mass will change the critical density of the universe. Assuming the transition leads to a smaller Planck mass (i.e., $\Omega_0 < 0$) and a higher $H^2$, will lead to a larger critical density of the universe. However, there is no change to physical matter density or physical radiation density during this transition, even if the fractional densities are reduced because the critical density increases. Without the introduction of a new energy component to compensate for this drop in the sum of the fractional energy densities of the universe, the curvature of the universe would have to change.

For the TPM model, the difference is made up by the scalar field that is coupled to gravity. Therefore, the equation of state parameter for the scalar field will tend to track the behavior of the dominant energy component at that particular time. This behavior manifests itself by a peak in the plot of $w_{TPM}(a)$ at a value above 0 when the universe is radiation dominated, switching to matter domination, and then settling to a value close to 0 during matter domination, as seen in Figure 1.

Considering linear scalar perturbations around a Friedmann–Lemaître–Robertson–Walker (FLRW) background in the Newtonian gauge, the Newtonian potential, $\Phi \equiv \delta g_{00}/(2g_{00})$, and the intrinsic spatial curvature, $\Psi \equiv -\delta g_{ii}/(2g_{ii})$, can be related to gauge-invariant comoving matter density fluctuations $\Delta_m$ through a modified Poisson equation and a modified lensing equation

$$\mu(k, a) \equiv \frac{-k^2 \Phi}{4\pi G a^2 \bar{\rho}_m \Delta_m}, \quad \Sigma(k, a) \equiv \frac{-k^2(\Phi + \Psi)}{8\pi G a^2 \bar{\rho}_m \Delta_m}, \quad (24)$$

where $\mu$ is called the effective gravitational coupling and encodes the effective modification induced in the Poisson equation. The $\Sigma$ parameter, called the light deflection parameter, describes the modifications of gravity on null geodesics, i.e., how the equation for the lensing potential $(\Phi + \Psi)/2$ is altered compared to standard general relativity. A nonparametric data-driven reconstruction of these functions has been presented in Raveri et al. (2021); Pogosian et al. (2021). A phenomenological transition in $\mu$ and $\Sigma$ at low redshift ($z_t < 4$) has been tested against cosmological data in Khosravi & Farhang (2022), finding no preference for such a modification of the cosmological model. However, a direct comparison with the TPM model is not possible due to the different epochs of transitions considered. Adopting the quasi-static approximation, valid for scales well inside the cosmological horizon, the expression for $\mu$ and $\Sigma$ in terms of EFT functions are given in Pogosian & Silvestri (2016). For the TPM model we find

$$\mu(k, a) = \frac{1}{1 + \Omega} \frac{1 + M_C^2 \frac{a^2}{k^2}}{f_3/(2f_1(1 + \Omega) m_0^2) + M_C^2 \frac{a^2}{k^2}}, \quad (25)$$

$$\Sigma(k, a) = \frac{1}{2(1 + \Omega)} \frac{1 + f_5/f_1 + 2M_C^2 \frac{a^2}{k^2}}{f_3/(2f_1(1 + \Omega) m_0^2) + M_C^2 \frac{a^2}{k^2}}, \quad (26)$$

Table 2
Maximum Likelihood (ML) Parameter Values and, in Parentheses, Mean plus 68% Confidence Level (CL) Bounds, for the Λ CDM, TPM f(R), and TPM Models, Using Our Baseline Data Set

| | Fit To Baseline Likelihood | | |
|---|---|---|---|
| Model | ΛCDM | TPM f(R) | TPM |
| $100\theta_{MC}$ | 1.04107 (1.04099 ± 0.00029) | 1.04116 ($1.04122^{+0.00032}_{-0.00037}$) | 1.04111 (1.04114 ± 0.00034) |
| $\Omega_b h^2$ | 0.02242 (0.02241 ± 0.00013) | 0.02243 (0.02249 ± 0.00015) | 0.02247 (0.02245 ± 0.00014) |
| $\Omega_c h^2$ | 0.11919 (0.11933 ± 0.00091) | 0.11896 (0.1189 ± 0.0010) | 0.1185 (0.1195 ± 0.0011) |
| $\tau$ | 0.0592 (0.0574 ± 0.0073) | 0.0575 (0.0566 ± 0.0077) | 0.0484 (0.0528 ± 0.0075) |
| $\ln(10^{10} A_s)$ | 3.052 (3.049 ± 0.014) | 3.051 (3.048 ± 0.016) | 3.029 (3.041 ± 0.015) |
| $n_s$ | 0.9677 (0.9666 ± 0.0036) | 0.9692 (0.9704 ± 0.0046) | 0.9681 ($0.9685^{+0.0041}_{-0.0046}$) |
| $\Omega_0$ | | −0.014 ($-0.0153^{+0.015}_{-0.0037}$) | −0.027 ($-0.0140^{+0.014}_{-0.0031}$) |
| $x_T$ | ... | −4.09 ($-4.90^{+1.1}_{-0.66}$) | −3.59 ($-5.05^{+1.1}_{-0.87}$) |
| $x_T$ (95% CL) | | $-4.9^{+1.5}_{-1.9}$ | $-5.1^{+1.5}_{-1.9}$ |
| $\sigma$ | ... | 0.873 ($1.50^{+0.66}_{-0.97}$) | 2.71 (<1.85) |
| $c_0$ | ... | 0 (fixed) | −0.0216 ($-0.0119^{+0.0046}_{-0.0064}$) |
| $c_0$ (95% CL) | | | $-0.0119^{+0.0099}_{-0.0090}$ |
| $w_{TPM,0}$ | ... | −0.9945 ($-0.9932^{+0.0037}_{-0.0083}$) | −1.05 ($-1.028^{+0.017}_{-0.020}$) |
| $H_0$ | 67.72 (67.65 ± 0.41) | 68.27 ($68.43^{+0.51}_{-0.83}$) | 69.19 ($69.22^{+0.67}_{-0.86}$) |
| $\sigma_8$ | 0.812 (0.8110 ± 0.0060) | 0.850 ($0.873^{+0.023}_{-0.071}$) | 0.9099 ($0.8530^{+0.0051}_{-0.040}$) |
| $S_8$ | 0.826 (0.826 ± 0.010) | 0.857 ($0.877^{+0.046}_{-0.078}$) | 0.903 ($0.8494^{+0.0089}_{-0.043}$) |
| $\Delta\chi^2_{CMB}$ | 0 | 0.5 | −5.88 |
| $\Delta\chi^2_{CMB\ lensing}$ | 0 | −0.05 | −0.05 |
| $\Delta\chi^2_{BAO}$ | 0 | −0.23 | 0.77 |
| $\Delta\chi^2_{Pantheon}$ | 0 | −0.05 | 0.26 |
| $\Delta\chi^2_{tot}$ | 0 | 0.18 | −4.8 |
| $\Delta\log_{10} B$ | 0 | −3.60 | −3.85 |

**Note.** Mean plus 95% CL bounds are also shown for selected parameters.

with

$$M_C^2 = \frac{3}{f_1}\left[m_0^2\left(2H\dot{H} + \frac{\ddot{H}}{2}\right)\dot{\Omega} + c\left(2\dot{H} - \left(\frac{\dot{H}}{H}\right)^2 + \frac{\dot{H}}{H}\right)\right] \quad (27)$$

$$f_1 = m_0^2 \frac{\dot{\Omega}^2}{1+\Omega} + c\frac{\dot{H}}{H^2} \quad (28)$$

$$f_3 = \frac{3}{2}(m_0^2\dot{\Omega})^2 + m_0^2\dot{\Omega}\frac{2c}{H} + m_0^2(1+\Omega)\frac{2c\dot{H}}{H^2} - 2\left(\frac{c}{H}\right)^2 \quad (29)$$

$$f_5 = m_0^2 \frac{\dot{\Omega}^2}{2(1+\Omega)} + c\frac{\dot{H}}{H^2} \quad (30)$$

where $\Omega$ and $c$ are defined in Equations (2) and (4), and the overdot represents derivatives with respect to time. The Compton mass of the scalar field, $M_C$, sets a scale, which makes the evolution of perturbations scale dependent. This scale dependence leads to a characteristic modification of the matter power spectrum, shown in Figure 2. On large scales compared to the Compton mass $k/a \ll M_C$, the effect of modified gravity on matter perturbations is small, and $\mu \sim \mu_0 = (1+\Omega)^{-1}$, $\Sigma \sim \Sigma_0 = (1+\Omega)^{-1}$, recovering the background value. In the opposite regime, on sub-Compton scales, $k/a \gg M_C$, matter perturbations feel the effect of an increased $\mu$. Indeed taking the limit $k \to \infty$ in Equation (25) we find $\mu_\infty = 2f_1/f_3$, $\Sigma_\infty = (1 + f_5/f_1)\mu_\infty$.[5] These quantities are shown in the first two panels of Figure 2. Deviations of the $\mu$ and $\Sigma$ functions from a value of 1 are the effects of modified gravity.

At early epochs, approximately in the four decades of scale factor centered on $a_T$, the main effect on $\mu$ and $\Sigma$ is sourced by the evolution of the coupling function $\Omega$. Indeed, we have $\dot{\Omega} \gg c/H$. In this regime the TPM model behaves like an $f(R)$ model, with $\mu_\infty \simeq 4/3(1+\Omega)^{-1}$, $\Sigma_\infty \simeq (1+\Omega)^{-1}$, and $aM_C \gg H$. The modes that have entered the horizon early enough to become sub-Compton during this regime, satisfying $k \gg aM_C \gg H$, experience a stronger gravity relative to larger wavelength modes, which leads to enhanced clustering on small scales. This effect manifests as a high-k bump in $\Delta P(k)/P_{\Lambda CDM}$.

At later epochs, when the effect of the $c_0$ parameter becomes dominant, i.e., $\dot{\Omega} \ll c/H$, the TPM model deviates from the $f(R)$ limit, showing a smaller $\mu_\infty \sim (1 + \Omega - c/\dot{H})^{-1}$ and a very small Compton mass $M_C \sim H$. This change in $\mu_\infty$ is clearly visible in Figure 2, close to $a = 10^{-3}$ for the best-fit TPM model. Conversely, $\Sigma_\infty$ remains constant in this phase, because the change in $\mu_\infty$ is compensated by an opposite change in the factor $1 + f_5/f_1$. Modes entering the horizon during this phase can be considered sub-Compton modes, and correspond to $k \lesssim 0.1\,\mathrm{h\,Mpc}^{-1}$. We see that for these modes the amplitude of $P(k)$ is very close to the $\Lambda$ CDM one, as these modes experience a lower deviation of $\mu$ from the general relativity value. The switch between the f(R) and the late-time limits in the TPM model depends on the $x_T$ and $\sigma$ parameters, which ends up influencing the amplitude of $P(k)$ at small scales. This, in turn, affects the parameter $S_8$. Later transitions lead to longer wavelength modes becoming sub-Compton while the TPM model is in the f(R) regime, which results in a larger value of $S_8$.

---
[5] We adopt the $\mu_\infty$ notation as it is used in other works. It should be noted that $\mu_\infty$ does not refer to solar system physics because nonlinear physics is not taken into account in this computation. In practice, the $\infty$ just means in the very large $k$ limit relative to the Compton mass.

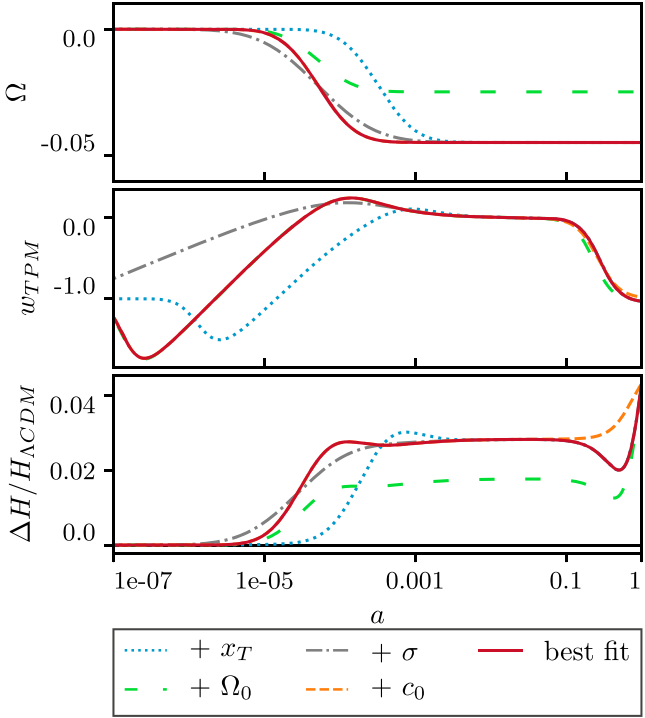

**Figure 1.** The evolution of the $\Omega$ (top), $w_{\rm TPM}$ (middle), and $H$ (bottom) as functions of the scale factor. For the Hubble rate, $H(a)$, we show relative deviations with respect to the best-fit $\Lambda$ CDM solution. Solid lines show theoretical functions computed using best-fit parameters from the fit to the Baseline + $H_0$ likelihood. Loosely dashed, dotted, and dashed–dotted lines are obtained by increasing the parameters $\Omega_0$, $x_T$, and $\sigma$ respectively by the amount of one standard deviation. For the $+c_0$ line we use $c_0 = -10^{-4}$ to better visualize the change. Varying $x_T$ changes the epoch of the transition of the $\Omega$ function at early times, which correspondingly affects the remaining functions, as discussed in the text. The parameter $\sigma$ sets how fast the evolution of $\Omega$, $w_{\rm TPM}$, and $H(a)$ is around the epoch of the transition. The parameter $\Omega$ modifies the background expansion throughout the entire cosmic history. The parameter $c_0$ modifies the expansion history at late times, for $a > 0.1$.

At late time, for $a \sim 1$, $\mu_\infty$ and $\Sigma_\infty$ grow again due to the contribution of $c/\dot{H}$. Values of $c_0$ close to 0 lead to $\mu_\infty = \Sigma_\infty = (1 + \Omega_0)^{-1}$. The late-time growth in $\mu_\infty$ determined by $c_0$ only produces a very small effect in the low-k region of $P(k)$.

### 4.2. Parameter Constraints from MCMC

The best-fit model residuals with $\Lambda$ CDM for the CMB TT, TE, EE, and $\phi\phi$ power spectra are shown in Figure 3. For the fit to the Baseline + $H_0$ likelihood, we find that overall, both the TPM and TPM f(R) models lead to an improvement over $\Lambda$ CDM with reductions in $\chi^2$ given by $\Delta\chi^2 = -23.72$ and $\Delta\chi^2 = -16.68$ respectively. The majority of this improvement comes from a better fit to the $H_0$ prior, as evidenced by the reductions in $\chi^2$ given by $\Delta\chi^2_{H_0} = -20.06$ and $\Delta\chi^2_{H_0} = -17.8$. While the TPM model is able to fit the CMB anisotropy data better than $\Lambda$ CDM with $\Delta\chi^2 = -6.29$, the TPM f(R) model fits slightly worse with $\Delta\chi^2 = 1.8$. For the TPM model, the fits to BAO and SN Ia are slightly worse, which is comparable to EDE models (e.g., Hill et al. 2020).

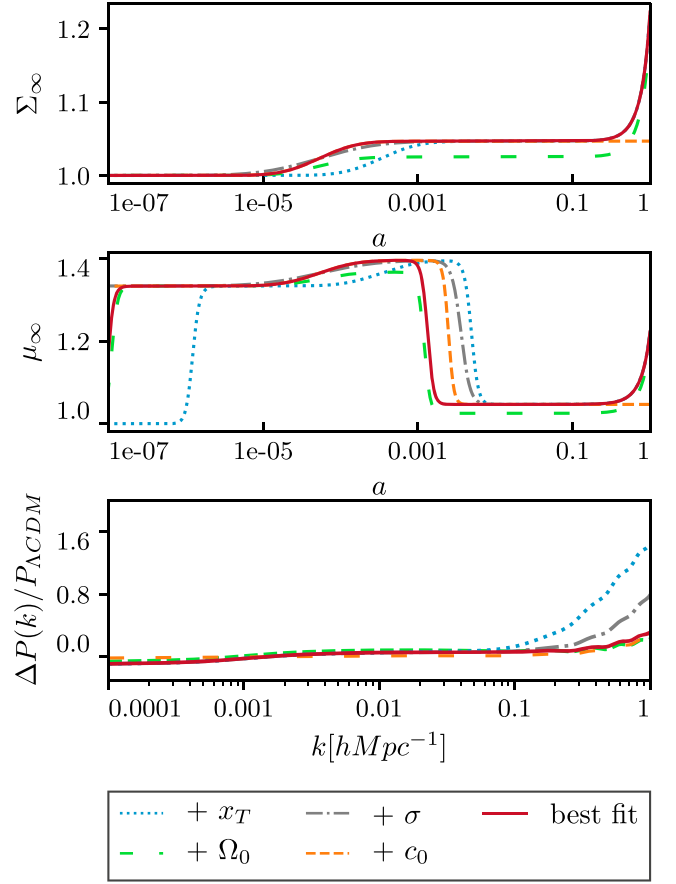

**Figure 2.** The effect of a variation of the TPM parameters on the $\Sigma_\infty$ (top) and $\mu_\infty$ (middle) functions, which quantify deviations from general relativity. We show relative deviations between the matter power spectra computed for the TPM and $\Lambda$ CDM models (bottom). Different curves are obtained by varying the TPM parameters as in Figure 1. With $\Omega$ closer to 0, as in the $+\Omega_0$ case, the model shifts toward $\Lambda$ CDM, reducing the deviation at small scales. The parameters $\sigma$ and $x_T$ modify the duration and epoch of transition, which affects the switch between higher f(R)-like values of $\mu_\infty$ and lower late-time values of this function, happening around $a = 0.002$ for the best-fit TPM model. Modes that enter the sub-Compton regime when $\mu_\infty$ is close to the f(R) value $\sim 4/3$ experience an enhanced clustering relative to $\Lambda$ CDM, leading to a boost in the value of $\Delta P(k)/P_{\Lambda CDM}$ at large $k$. As we see, a bigger $x_T$ and $\sigma$ boost the $P(k)$ on small scales. The parameter $c_0$ mainly affects the $a \sim 1$ evolution of $\mu_\infty$ and $\Sigma_\infty$, with a small impact on the $P(k)$.

Excluding the prior on $H_0$ from local measurements, the overall reduction in $\chi^2$ is noticeably smaller for both the TPM and TPM f(R) models. While the TPM model does fit the CMB data better than $\Lambda$ CDM, the reduction in $\chi^2$ is not large enough to justify the addition of four new model parameters.

This can be quantified by the evidence factor $\Delta\log_{10} B$ shown in Table 2. Negative values of the evidence factor show that the TPM model is disfavored over $\Lambda$ CDM by our Baseline data set, due to the increased number of free parameters, despite showing comparable (TPM f(R)) or better (TPM) values for $\chi^2$ in the best-fit case. Nevertheless, the $H_0$ tension is still reduced to a value of $\sim 2.8\sigma$ w.r.t. $\Lambda$ CDM. Conversely the evidence factor strongly favors TPM when local measurements of $H_0$ are included in the analysis, as shown in Table 1.

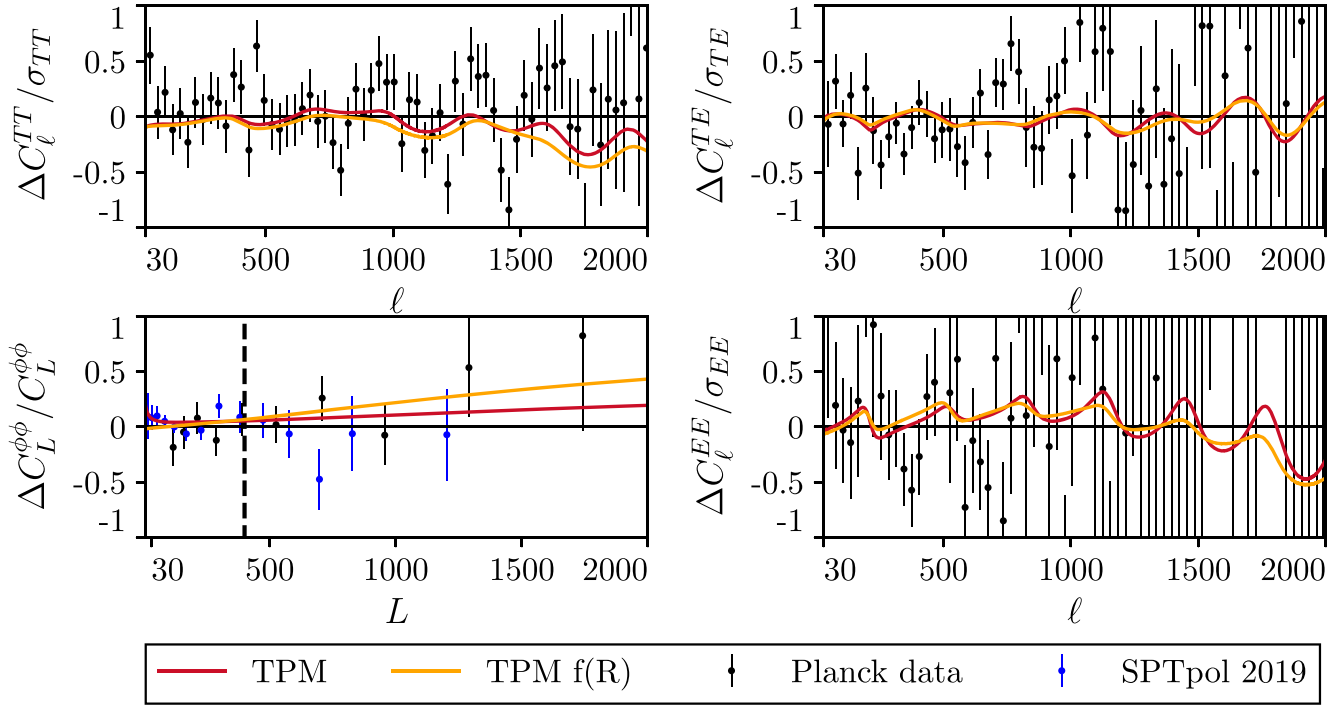

**Figure 3.** Plots of the differences, in units of cosmic variance, between the best-fit TPM (red) or TPM f(R) (orange) and the best-fit $\Lambda$ CDM models where the best-fit model parameters are derived from a maximum likelihood test using CMB, BAO, and SN Ia data along with an $H_0$ prior from local measurements. CMB data is remarkably constraining in both the TPM and TPM f(R) models. Both models prefer a small-scale enhancement of the lensing power. In the lower-left panel, we include a vertical dashed line to represent the upper limit of the lensing modes included in the suggested Planck lensing likelihood. We find the results are robust to using either the aggressive Planck lensing likelihood or the SPTpol 2019 lensing likelihood.

Given the residual level of inconsistency between local estimates of $H_0$ and early-time cosmological data, we caution the reader against an overoptimistic interpretation of the results obtained for TPM in the Baseline + $H_0$ configuration. However, as discussed below and shown in Figure 4, the results obtained in this configuration clearly show the existence of interesting degeneracy directions in the parameter space of TPM, that can be exploited to bring late- and early-time cosmological measurements in better agreement, and reduce the $S_8$ and $H_0$ tension at once. This interesting feature of the TPM scenario shows how an early-time variation of the Planck mass produces inherently different results and has the potential of accommodating data better than a simple variation of the equation of state.

In Figure 4 we show marginalized joint posterior distributions for a set of TPM and TPM f(R) independent model parameters together with the derived parameter $S_8$ when fit to the Baseline + $H_0$ likelihood. The TPM model prefers an increase in the best-fit value of $H_0$ to $H_0 = 71.09 \pm 0.75$ kms$^{-1}$ Mpc$^{-1}$ when fit to the Baseline + $H_0$ likelihood and $H_0 = 69.22^{+0.67}_{-0.86}$ kms$^{-1}$ Mpc$^{-1}$ when the $H_0$ prior is excluded. These correspond to 1.2$\sigma$ and 2.8$\sigma$, respectively, below the SH0ES constraint.

For the TPM fit to the Baseline + $H_0$ likelihood, the amount of deviation of $\Omega$ from 0 is constrained to be within 10% at the 2$\sigma$ level. In particular, we find $\Omega_0 = -0.050 \pm 0.019$ for the TPM model and $\Omega_0 = -0.065 \pm 0.018$ for the TPM f(R) model when the $H_0$ prior is included. Excluding the $H_0$ prior, the preference for $\Omega_0 < 0$ falls to 1$\sigma$ for the TPM case. This results from the data not requiring higher $H_0$ values that could be achieved by larger amplitude transitions in $\Omega$.

Including the $H_0$ prior in the fit with the TPM model, the duration of the transition, measured by the $\sigma$ parameter, is constrained to be lower than 2.1 e-folds at the 95% confidence level. The transition is also constrained to happen during the radiation-dominated epoch. The constraint on $x_T \equiv \log_{10}(a_T)$ is $x_T = -5.32^{+0.96}_{-0.72}$ meaning the transition in the Planck mass can happen over multiple decades of scale factor growth prior to recombination and still lead to a better fit to SH0ES data today. We find similar behavior for the TPM f(R) model. Excluding the prior on $H_0$ from local measurements, there is still a preference for the transition to happen during the radiation-dominated epoch with $x_T = -5.05^{+1.1}_{-0.87}$.

For the TPM model, where $c \neq 0$, the late-time effect of a negative $c_0$ parameter also contributes to bring the present $H_0$ to higher values, through a reduction of the present value of $w_{\text{TPM},0}$ to below $-1$ for negative values of $c_0$, as seen in Equation (23). In the TPM f(R) model, $c_0 = 0$, so to accommodate the $H_0$ prior, it is necessary to have a larger shift in the $\Omega_0$ parameter compared to the TPM case. This leads to a slightly worse fit to Planck CMB anisotropy data, as evidenced by the $\Delta\chi^2 = 1.8$ for TPM f(R) and $\Delta\chi^2 = -6.29$ for TPM when both are fit to the Baseline + $H_0$ likelihood. Similarly, most of the increase in $H_0$ relative to the $\Lambda$ CDM case for the TPM model fit when the $H_0$ prior is excluded comes from the downward shift in the value of $c_0$ as evidenced by the lack of an increase in $H_0$ in the TPM f(R) case where $c_0$ is fixed to 0.

The enhanced matter power spectrum in the high-$k$ ($\gtrsim 10^{-1}h$ Mpc$^{-1}$) region generally leads to an increased preferred value of the $\sigma_8$ parameter for the TPM model relative to Λ CDM when fit to the Baseline + $H_0$ likelihood, seen in Table 1. Similarly, there is also a slight increase in the preferred value of $n_s$ at about the 1σ level in the TPM model fit to Baseline + $H_0$ likelihood compared to Λ CDM. This shift is comparable to similar shifts in the value of $n_s$ preferred by data when fit with EDE models.

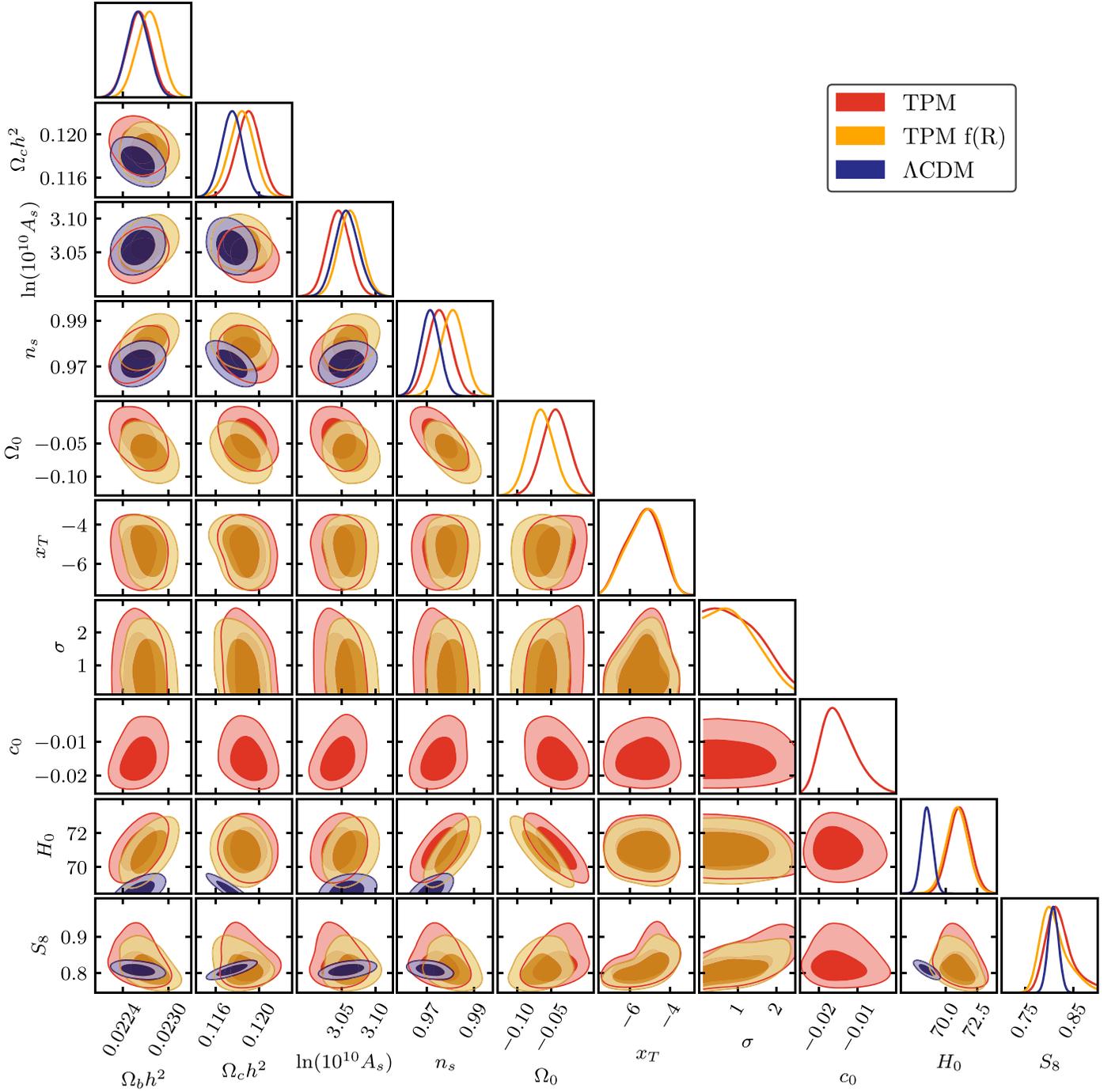

**Figure 4.** The marginalized joint posterior of the parameters of the TPM model, obtained using our combined data sets. ΛCDM results are shown for comparison. The Baseline + $H_0$ likelihood prefers approximately a 5% shift downward in the parameter $\Omega_0$, which corresponds to a roughly 5% upward shift in the Planck mass on cosmological scales. This shift results in an increase in the preferred value of the Hubble constant to $H_0 = 71.09 \pm 0.75$ kms$^{-1}$Mpc$^{-1}$. In the case where the $H_0$ prior is excluded, we still find an increase in the Hubble constant relative to ΛCDM to $H_0 = 69.22^{+0.67}_{-0.86}$ kms$^{-1}$Mpc$^{-1}$ (not shown). Additionally, there is a sizable part of the $H_0$ and $S_8$ parameters where $H_0 > 70$ kms$^{-1}$Mpc$^{-1}$ and $S_8 < 0.8$. This parameter space is available to the TPM model because of variations in the matter fraction $\Omega_m$ as shown in Figure 5. Finally, the transition in the value of $\Omega_0$ can occur at any time over the two decades in scale factor growth preceding matter–radiation equality, as shown by the posterior for $x_T$.

We now consider the impact of TPM model on the combination $S_8 \equiv \sigma_8 \sqrt{\Omega_m/0.3}$, constrained by weak lensing surveys. For the TPM and TPM f(R) models, the marginalized constraints on $\Omega_m h^2$ are only slightly shifted with respect to ΛCDM, as shown in Figure 4 and Table 1. This is in contrast with EDE-like models, in which the $\Omega_m h^2$ tends to be remarkably shifted to higher values to better accommodate CMB and SN Ia data. In Vagnozzi (2021), it is suggested that the early dark energy leads to an enhanced early Integrated Sachs–Wolfe (ISW) effect, which can be compensated for by an increase in the physical cold dark matter density. For the TPM model, the scalar field during matter domination behaves like matter, with $w_{\rm TPM} \simeq 0$, thus compensating for the enhanced early ISW effect. This is visible in Figure 3, which shows that the CMB TT power spectrum obtained with the best-fit $\Lambda$ CDM and TPM models have roughly the same amplitude around the scale of the first acoustic peak despite no substantial increase in the preferred value of $\Omega_c h^2$.

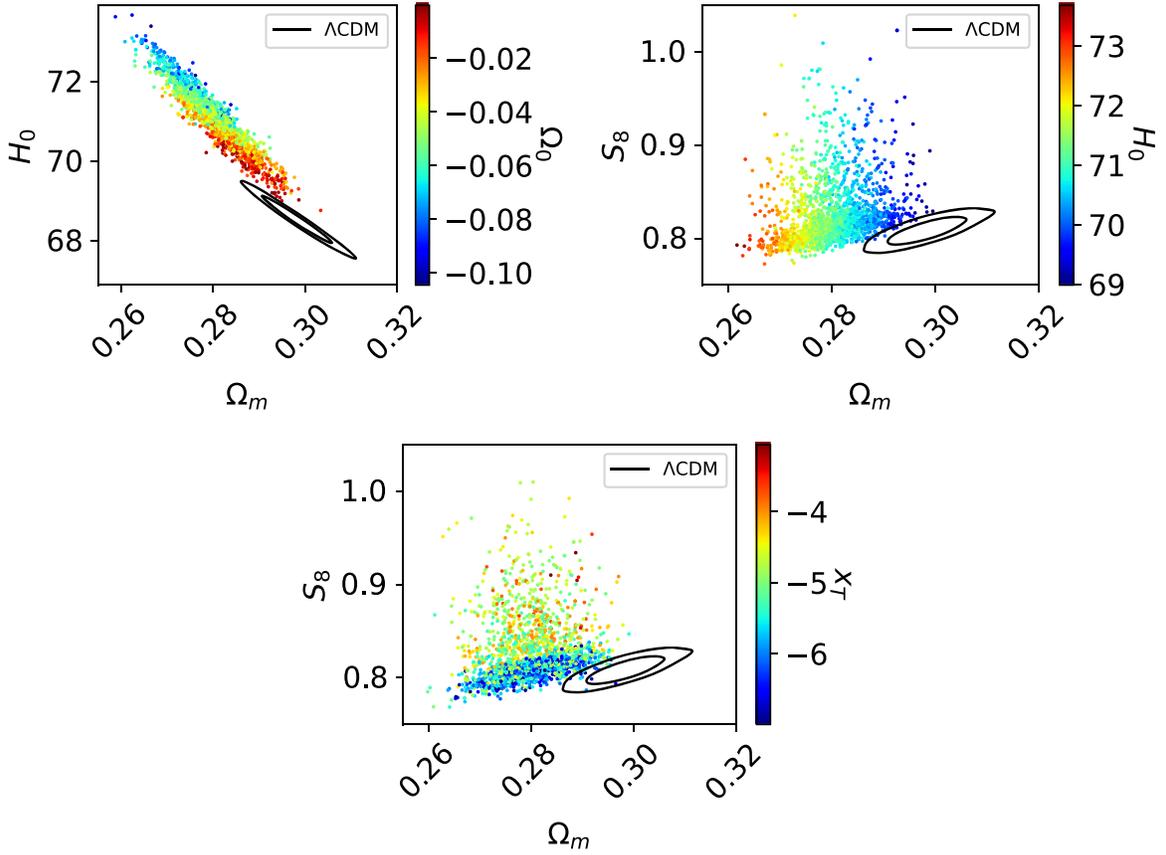

**Figure 5.** Plots of the posterior distributions for the TPM model fit to the Baseline + $H_0$ likelihood. Additionally, we add black contours to represent the $\Lambda$ CDM fit to the Baseline + $H_0$ likelihood. The top left panel shows the relationship between $H_0$, $\Omega_m$, and $\Omega_0$. In particular, decreasing the value of $\Omega_0$, which sets the amplitude of the change of the effective Planck mass on cosmological scales, results in a shift along the $\Omega_m - H_0$ degeneracy toward higher values of $H_0$ and lower values of $\Omega_m$. The top right panel shows that higher values of $H_0$ tend to correspond to lower values of $S_8$ primarily through the reduction of $\Omega_m$. The bottom panel shows that the larger values of $S_8$ (those above ~0.83) correspond to parts of the parameter space where the transition in $\Omega$ happens after $x_T \equiv \log_{10}(a_T) = -6$. This results from longer wavelength modes experiencing a longer period of enhanced gravity relative to $\Lambda$ CDM, which results in increased clustering on small scales.

Additionally, in minimally coupled and flat cosmological models like EDE, the Pantheon SN Ia data directly constrain $\Omega_m$, which results in an increase in $\Omega_m h^2$ when $H_0$ increases. The higher physical matter density results in an increase in the amount of clustering of matter, which in turn leads to a higher $S_8$ value when the $H_0$ prior from local measurements is included in the analysis. In the TPM and TPM f(R) model, this effect is countered by the fact that the scalar-field density gives a non-negligible contribution to the energy balance of the universe after the transition, which determines an increased critical density and leads to a decrease in $\Omega_m$ constrained by SN Ia data. This leads to a reduction in the $S_8$ parameter relative to $\Lambda$ CDM. This degeneracy between $H_0$ and $\Omega_m$ for both the TPM and $\Lambda$ CDM models is shown in Figure 5. For the TPM model, $\Omega_0$ approximately sets the location along this degeneracy of the $H_0$ and $\Omega_m$ parameter values. This degeneracy is not exact because the $c_0$ parameter also affects the value of $H_0$.

Figure 5 also shows that higher values of $H_0$ tend to prefer lower values of both $\Omega_m$ and $S_8$. This degeneracy is strongest with early, rapid transitions in $\Omega$, as later, slower transitions allow for longer wavelength modes to experience modified gravity for a longer period of time, as discussed in Section 4.1. This enhances the clustering of matter at scales $k \sim 0.1$ $h\,{\rm Mpc}^{-1}$, which increases $\sigma_8$ and thus $S_8$.

Figure 6 shows the effect of TPM and TPM f(R) models on the characteristic scales related to the physics of the acoustic peaks in the CMB power spectrum. In all cases, the posteriors are for the fit to the Baseline + $H_0$ likelihood. Because the transition occurs well within the radiation-dominated epoch, there is a rescaling of $H(a)$ that roughly preserves the shape of the $\Lambda$ CDM model from well before the recombination epoch to very late times. As discussed in Knox & Millea 2020, this allows for a reduction of the comoving sound horizon to last scattering, $r_*$, while preserving the characteristic angular scale of matter-radiation equality, $\theta_{\rm eq}$, and the shape of the radiation-driving envelope, which helps to reduce the $H_0$ tension.

The preferred value of $r_*$ in the TPM and TPM f(R) models is lower than $\Lambda$ CDM by 2.4 and 3.5 $\sigma$ respectively, in the Baseline + $H_0$ configuration. Because the TPM model also allows a late-time enhancement of the expansion rate from the $c_0$ parameter, the reduction in $r_*$ tends to be smaller than in the TPM f(R) case. The reduction in the size of the sound horizon must be compensated for by a similar reduction in the size of the angular diameter distance to the sound horizon, $D_{A*}$, to maintain the value of $\theta_* = r_*/D_{A*}$. Because the angle of the characteristic scale of photon diffusion damping, $\theta_D$, has the same angular diameter distance to the last scattering, it follows that there must be some change to either the angle, $\theta_D$, or the characteristic scale of photon diffusion at last scattering. In practice, there is a compromise between them that results in an enhanced photon diffusion damping. As constraints on the angle $\theta_D$ become more precise, the TPM model will tend to predict a larger photon diffusion damping effect, which could be used to distinguish between the TPM and other models.

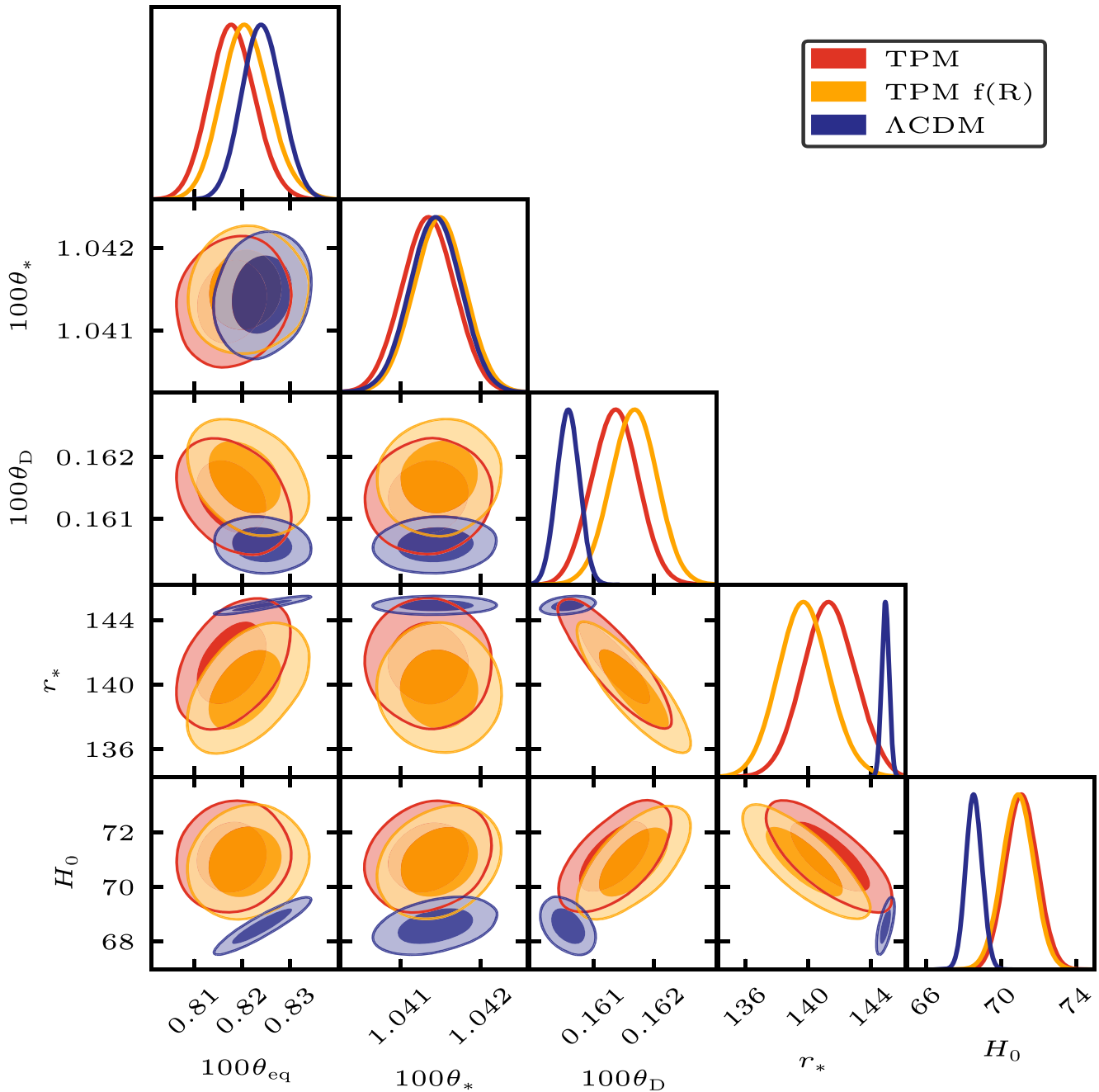

**Figure 6.** Constraints on characteristic scales related to the physics of the CMB using the Baseline + $H_0$ likelihood. These characteristic scales are the angles associated with the comoving size of the sound horizon at matter-radiation equality, $\theta_{\rm eq}$, the comoving size of the sound horizon at last scattering, $\theta_*$, and the comoving size of the photon diffusion scale at last scattering, $\theta_D$. Much like in EDE models, the TPM and TPM f(R) allow for a larger value of $H_0$ by reducing the comoving size of the sound horizon, $r_*$ relative to $\Lambda$ CDM. For the TPM model, this reduction does not need to be as large as the TPM f(R) case because varying the $c_0$ parameter allows for $w_{\rm TPM,0} < -1$, which allows for an increase in $H_0$. For the TPM and TPM f(R) models, there is a weak degeneracy between $H_0$ and $\theta_{\rm eq}$. Because $r_*$ is reduced and the angular diameter distance to $\theta_*$ and $\theta_D$ is the same, it follows that if $\theta_*$ must remain fixed, then $\theta_D$ must shift. This shift results in a strong positive degeneracy between $H_0$ and $\theta_D$ meaning larger values of $H_0$ correspond to an increase in photon damping.

## 5. Conclusions

Motivated by the parameter tensions in $\Lambda$ CDM, we have investigated the possibility that this constitutes a failure of general relativity. In this process we do not invoke a nonzero cosmological constant, potentially rendering vacuum fluctuations technically natural.

We have constructed a modified gravity model that allows for a phenomenological shift in the effective Planck mass on cosmological scales prior to recombination through a non-minimal coupling of a scalar field with gravity. We called this model the Transitional Planck Mass or TPM model. To construct the TPM model, we used the Effective Field Theory of Dark Energy and Modified Gravity (EFT) formalism and the following assumptions and definitions:

1. The universe is flat at all times.
2. The TPM model is a modified gravity model that gives second-order equations of motion for the scalar field and therefore can be mapped into the class of Horndeski models.
3. The TPM model includes a scalar field that is nonminimally coupled to gravity that causes the transition in the value of the effective Planck mass on cosmological scales.
4. The speed of propagation of gravitational waves is equivalent to the speed of light.
5. The transition in the value of the Planck mass on cosmological scales is quantified by the EFT $\Omega$ function. We choose the transition to be a single step represented by a simple error function and not some more complicated description.
6. The effective value of the Planck mass on cosmological scales today does not equal the solar system value. There must be some screening mechanism to account for the differences in values.
7. We normalized the initial value of the $\Omega$ function such that the effective Planck mass on cosmological scales prior to the transition does equal the value measured locally today.
8. The EFT functions $\{\bar{M}_2^2, \bar{M}_3^2, m_2^2, \hat{M}^2\}$ are all fixed to 0. This is required by the previous assumptions.
9. The EFT $c$ function, which along with the $\Lambda$ function sets the behavior of the scalar field that is coupled to gravity, is constant with time. The remaining functions $M_2^4$ and $\bar{M}_1^3$ are fixed so that their contribution to cosmological perturbations is of the same order as the one given by the $c$ function.
10. The model should be free of ghost and gradient instabilities.

It is possible to construct modified gravity models that have many additional parameters that are largely unconstrained by cosmological measurements and thus may be capable of sufficiently degrading the precision of parameter constraints to allow for a complete resolution to cosmological parameter tensions. In contrast to this concern, we find that the TPM model is precisely constrained by CMB + BAO + SN Ia data along with a prior from local measurements of $H_0$. Moreover, when the $H_0$ prior is not included, the $H_0$ tension is reduced to $\sim 2.8\sigma$ with an increase to $H_0 = 69.22^{+0.67}_{-0.86}$ kms$^{-1}$Mpc$^{-1}$.

The TPM model fit to the Baseline + $H_0$ likelihood shows a preference for about a 5% shift upward in the effective Newton's constant on cosmological scales today. Additionally, we find that there is a 2.5$\sigma$ preference for a nonzero transition ( i.e., $\Omega_0$ is 2.5$\sigma$ from 0). This results in a preference for a higher Hubble constant that is within 1$\sigma$ of the $H_0$ measurement. There is a significant improvement in the $\chi^2$ fit to the Baseline + $H_0$ data ($\Delta\chi^2 = -23.72$ with respect to $\Lambda$ CDM), with the primary reduction in $\chi^2$ coming from the $H_0$ prior ($\Delta\chi^2 = -20.06$ with respect to $\Lambda$ CDM). This is evidenced by both the increase in the preferred value of $H_0$ and the milder improvement in the fit to the Baseline likelihood compared to $\Lambda$ CDM when the $H_0$ prior is excluded. The TPM model is able to simultaneously fit the CMB anisotropy and lensing data better than $\Lambda$ CDM, though not at a significant level given four additional model parameters.

The increase in the preferred value of $H_0$ when fitting the TPM model to the Baseline or Baseline + $H_0$ likelihoods is derived from two main effects. The primary effect is that changing the value of $\Omega_0$ allows for shifts along the degeneracy between $\Omega_m$ and $H_0$ present in CMB anisotropy data. Qualitatively, the data allow for a reduction in the matter fraction if correspondingly the gravitational attraction is increased. This degeneracy is preserved as long as flatness is assumed. Additionally, the slightly negative value of the $c_0$ parameter shifts the equation of state to be less than $-1$ today, leading to a late-time enhancement of the expansion rate.

While the mean value of $S_8$ for the TPM model fit to the Baseline + $H_0$ likelihood is higher than in the $\Lambda$ CDM case, the TPM model allows a region of parameter space where $S_8 < 0.8$ and $H_0 > 70$ kms$^{-1}$Mpc$^{-1}$. This suggests that the TPM model may also help relieve the clustering tension, although we caution here that a comparison with $S_8$ values derived from galaxy weak lensing surveys assuming $\Lambda$ CDM is not straightforward. Unlike EDE models, the TPM model modifies the late-time growth rate and lensing potentials (Figure 2). Additionally, the weak lensing measurements are sensitive to nonlinear scales. Analyzing how the TPM model performs with weak lensing data would require the implementation of code to handle the nonlinear effects, which we leave to future work.

An advantage of the TPM model is that its transition is free to occur over multiple decades in scale factor growth prior to recombination and thus does not lead to any coincidence problems. The best-fit value occurs at a scale factor $5.1 \times 10^{-5}$ when fit to the Baseline + $H_0$ likelihood. The TPM modified gravity model provides a dark energy-like effect in the late universe. This eliminates the need for a cosmological constant. The constraint on the $w_{\rm TPM}$ parameter in the late universe comes from the combination of CMB, BAO, and SN Ia data, which require a dark energy equation of state parameter near $-1$ today.

While this particular modified gravity model does not fully resolve the Hubble or clustering tensions without an $H_0$ prior to local measurements, we conclude that it may be possible to construct early universe modified gravity models that are able to fit current data better than $\Lambda$ CDM. We conclude that these new models can be precisely constrained even with present cosmological data. In the future, more precise measurements of the CMB damping tail, the CMB lensing power spectrum, and large-scale structure will distinguish between nonminimally coupled scalar-field models like the TPM model and $\Lambda$ CDM. This opens up a new direction for possible solutions to the Hubble and clustering tensions to be explored. These sorts of models are also computationally attractive to explore as there are publicly implemented codes such as EFTCAMB and EFTCosmoMC that can easily include these types of modifications. Finally, this model provides an alternative to $\Lambda$ CDM, which is valuable for future hypothesis testing when upcoming data become available.

This work was performed, in part, for the Jet Propulsion Laboratory, California Institute of Technology, sponsored by the United States Government under the Prime Contract 80NM00018D0004 between Caltech and NASA under subcontract numbers 1643587 and 1658514. We also acknowledge the NASA ADAP grant 80NSSC19K0526.

Calculations for this paper used computational resources at the Maryland Advanced Research Computing Center (MARCC). We acknowledge the use of the Legacy Archive for Microwave Background Data Analysis (LAMBDA), part of the High Energy Astrophysics Science Archive Center (HEASARC). HEASARC/LAMBDA is a service of the Astrophysics Science Division at the NASA Goddard Space Flight Center.

We thank Marco Raveri for his useful discussions.

**ORCID iDs**

Giampaolo Benevento  https://orcid.org/0000-0002-6999-2429
Joshua A. Kable  https://orcid.org/0000-0002-0516-6216
Graeme E. Addison  https://orcid.org/0000-0002-2147-2248
Charles L. Bennett  https://orcid.org/0000-0001-8839-7206